# Multi-Constrained 3D Topology Optimization Via Augmented Topological Level-Set

Shiguang Deng, Krishnan Suresh *

Mechanical Engineering, University of Wisconsin, Madison, USA

**Abstract**

The objective of this paper is to introduce and demonstrate a robust methodology for solving multi-constrained 3D topology optimization problems. The proposed methodology is a combination of the topological level-set formulation, augmented Lagrangian algorithm, and assembly-free deflated finite element analysis (FEA).

The salient features of the proposed method include: (1) it exploits the topological sensitivity fields that can be derived for a variety of constraints, (2) it rests on well-established augmented Lagrangian formulation to solve constrained problems, and (3) it overcomes the computational challenges by employing assembly-free deflated FEA. The proposed method is illustrated through several 3D numerical experiments.

## 1. INTRODUCTION

Over the last two decades, topology optimization (TO) [1] has accelerated from an academic exercise into an exciting discipline with, potentially, numerous industrial applications. The focus of this paper is specifically on *constrained* TO where several performance and manufacturing constraints must be considered during optimization.

In structural mechanics, a constrained TO problem may be posed as (see Figure 1):

$$\begin{aligned}&\underset{\Omega \subset D}{Min}\, \varphi(u,\Omega)\\ &g_i(u,\Omega) \leq 0; i=1,2,...,m\\ &\text{subject to}\\ &Ku = f\end{aligned} \quad (1.1)$$

where:

$$\begin{aligned}\varphi &: \text{Objective to be minimized}\\ \Omega &: \text{Topology to be computed}\\ D &: \text{Domain within which the topology must lie}\\ u &: \text{Finite element displacement field}\\ K &: \text{Finite element stiffness matrix}\\ f &: \text{External force vector}\\ g_i &: \text{Constraints}\\ m &: \text{Number of constraints}\end{aligned} \quad (1.2)$$

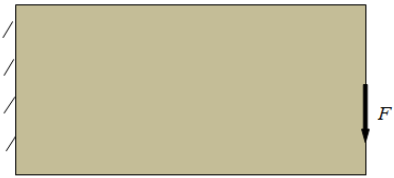

**Figure 1:** A single-load structural problem.

Various methods have been proposed to solve specific instances of Equation (1.1); these are reviewed in Section 2. For example, a

\* Corresponding author.
   E-mail address: ksuresh@wisc.edu

special case of Equation (1.1) is the *compliance-constrained* volume minimization problem:

$$\begin{aligned}&\underset{\Omega \subset D}{Min}\, |\Omega|\\ &J \leq J_{all}\\ &\text{subject to}\\ &Ku = f\end{aligned} \quad (1.3)$$

where:

$$\begin{aligned}J &: \text{Compliance}\\ J_{all} &: \text{Compliance allowable}\end{aligned} \quad (1.4)$$

Figure 2 illustrates the solution to a specific instance of Equation (1.3), where the allowable compliance is 60% larger than the initial compliance.

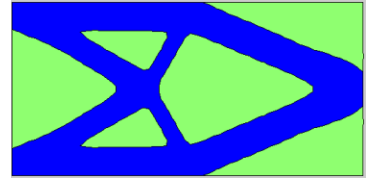

**Figure 2:** Optimal topology for a specific instance of Equation (1.3) over the structure in Figure 1.

In practice, additional constraints including stress, buckling, Eigen-value, and manufacturing constraints must be taken into account. The objective of this paper is to develop a unified method that can solve such multi-constrained TO problems. The proposed method and its implementation are discussed in Section 3. In Section 4, numerical experiments are presented, followed by conclusions in Section 5.

## 2. LITERATURE REVIEW

### 2.1 Constrained Topology Optimization

To solve a constrained TO problem, a TO formulation and a constrained optimization algorithm must be chosen.

Various *TO formulations* including homogenization [2], Solid Isotropic Material with Penalization (SIMP) [3] and level-set [4], [5], have been proposed. Constrained optimization algorithms, on the other hand, include method of moving asymptotes (MMA) [6], optimality criteria (OC) [7], simplex method [8], interior point method [9], Lagrangian multiplier method [9], augmented Lagrangian method [9] and so on.

We review below various combinations of TO formulations and optimization algorithms that have been proposed. Table 1 provides a chronological summary of relevant literature. The table and the review that follows are representative but not exhaustive; for example, constrained ground structures methods [10], [11], [12] are not reviewed here.

<u>SIMP Based Methods</u>

SIMP is perhaps the most popular TO formulation due to its simplicity, generality and success in several applications [13]. Based on the finite element method (FEM), SIMP assigns each element with a pseudo-density, and the pseudo-densities are then optimized to meet the desired objective [14].

Initially, SIMP was employed to solve compliance minimization problems [14]; it then evolved to include constraints. For example, one of the earliest SIMP-based stress-constrained TO implementation was reported in [15] where authors coalesced local stress constraints into a global stress constraint, and addressed instability issues via a weighted combination of compliance and global stress constraints. Further research on compliance and stress-constrained SIMP-based TO are discussed in [13], [16], [17], [18] and [19].

In [20], the authors proposed a SIMP-based trust-region method combined with augmented Lagrangian to solve a TO problem of continuum structures subject to failure constraints. In [21], a Heaviside design parameterization was used in SIMP to consider manufacturing constraints. The authors in [22] implemented SIMP with MMA to solve a TO problem with compliance and manufacturing constraints. In [23], using SIMP, a manufacturing constraint and a unilateral contact constraint were absorbed into compliance minimization formulation through augmented Lagrangian method. In [24], the authors used a modified SIMP formulation coupled with quadratic programming technique to minimize structural weight subject to multiple displacement constraints. The authors in [25] used MMA to solve a topology optimization problem with a probability-based high-cycle fatigue constraint. In [26], an algorithm was proposed to address multi-scale topology optimization problems subject to multiple material design constraints. In [27], a multi-point approximation algorithm was used as optimizer in a continuum structure topology optimization problem subject to dynamic constraints.

ESO/BESO Based Methods

ESO [28] is an alternate TO formulation where finite elements are gradually removed during each iteration. BESO [29] addresses some of the limitations of ESO by permitting the insertion of elements.

In [30], a principal-stress based ESO method was proposed to find the optimal design of cable-supported bridges subject to displacement and frequency constraints. During each optimization iteration, based on a threshold, elements were removed from the design domain. A similar method was used in [31] to solve contact design problems, where the authors proposed the interfacial gap between components be treated as optimization variables, while the contact stress be treated as an objective function. In [32], the Lagrangian multiplier method was used with BESO to combine the objective function of structural stiffness with a local displacement constraint. In [33], a modified BESO method was combined with optimality criteria to solve a topology optimization problem with natural frequency constraints. The authors argued this method can successfully avoid artificial local modes.

Level-Set Based Methods

Level-set formulation is gaining popularity in TO since it permits an unambiguous description of the boundary, and therefore permits imposition of constraints on the boundary. The level-set formulation relies on an evolving level-set which is typically controlled via Hamilton-Jacobi equations [34]. Readers are referred to [34] for a recent review of the success of level-set based methods in structural TO.

In [35], X-FEM based level-set and OC method were combined to find optimal designs for continuum structures with geometric constraints. In [36], a topological level-set method was coupled with an adapted weight method for solving stress-constrained compliance minimization problem. In [37], the authors combined classic shape derivative and level set method for front propagation; the Lagrangian multiplier technique was used for perimeter-control. Since there was no implemented mechanism for creation of holes, the final design was dependent on initial material layout. In [38], the augmented Lagrangian technique was combined with the topological sensitivity based level-set method to handle displacement, stress and compliance constraints.

In [39], level-set/X-FEM combined with a shape equilibrium constraint strategy was proposed. Specifically, a TO problem with stress constraint was formulated through Lagrangian multiplier method which was then iteratively solved. In [40], a level-set based method was derived to handle casting constraints; augmented Lagrangian method was applied for posing the constraints and calculating the shape derivative of objective function. In [41], a level-set based method was applied to the representative wing box of NASA Common Research Model to find the optimal 3-D aircraft wing structures. Compliance was minimized while balancing the aerodynamic lift and total weight. The level-set was shown to be robust and efficient by finding optimum solutions for multiple aerodynamic and body force load cases.

**2.2 Proposed Method**

From the above literature review, and from Table 1, one can conclude that significant progress has been made in recent years on constrained TO. Yet, a single method that can handle a variety of constraints, especially in 3D, has not been reported.

In this paper, we extend the 2D method proposed in [38] to achieve this goal; the main contributions of this paper are:

- The topological level-set in 3D is combined with the augmented Lagrangian method. The 3D computational challenges are addressed by exploiting the assembly-free deflated FEA [42].
- While only displacement and stress constraints were addressed in [38], additional buckling and Eigen-value constraints, are included here. Inclusion of buckling and Eigen-value constraints necessitates the need for soft and hard constraints, discussed later in the paper.
- Casting constraints are also addressed in this paper.
- While single-load problems were considered in [38], multiple and multi-load problems are considered here.

**3. TECHNICAL BACKGROUND**

In this section, we review several important concepts that contribute to the proposed work.

**3.1 Topological Sensitivity**

The proposed constrained TO method relies on the concept of topological sensitivity that is defined as the first order sensitivity of a quantity of interest with respect to infinitesimal change in topology. This was first explored by Eschenauer [43], and later extended by many other researchers [44], [45], including its generalization to arbitrary features [46], [47].

**Table 1:** Constrained topology optimization methods.

| Year | Authors | TOPO form. | Opt. solver | Manu. | Disp. | Comp. | Stress | Eigen | Buck. | Dim. |
|---|---|---|---|---|---|---|---|---|---|---|
| 1987 | Svanberg [6] | Convex approximation | MMA | | √ | | √ | √ | | 2 |
| 1992 | Zhou, Rozvany [48] | SIMP | OC | | | √ | | | | 2 |
| 1996 | R. Haber [49] | Penalized Homogenization | Interior penalty method | √ | | √ | | | | 2 |
| 1997 | M. Kocvara [10] | Ground structure approach | Interior point method | | √ | √ | | | | 2 |
| 1998 | J. Petersson, O. Sigmund [50] | SIMP | Linear programming | √ | | √ | | | | 2 |
| 2001 | L. Yin, et. al. [51] | SIMP | OC | √ | | √ | √ | | | 2 |
| 2002 | H. Guan, Y. Chen[30] | ESO | Parameterized criteria | | √ | | √ | √ | | 2 |
| 2003 | W. Li, Q. Li [31] | ESO | Parameterized criteria | √ | | | | | | 2 |
| 2004 | J. Pereira, E. Fancello [20] | SIMP | Augmented Lagrangian | | | | √ | | | 2 |
| 2004 | G. Allaire, et.al. [52] | Level-set | Lagrangian multiplier | √ | | √ | | | | 2,3 |
| 2006 | K. Zuo, L. Chen [22] | SIMP | Modified MMA | √ | | √ | | | | 3 |
| 2007 | M. Stolpe, T. Stidsen [53] | Hierarchical optimization | Linear programming | | √ | | √ | | | 2 |
| 2008 | M. Werme [17] | SIMP | Linear programming | | | √ | √ | | | 2 |
| 2008 | M. Bruggi, P. Venini [18] | SIMP | MMA | √ | | √ | √ | | | 2 |
| 2009 | J. Paris, F. Casteleiro [54] | SIMP | Simplex method | | | √ | √ | | | 2,3 |
| 2009 | A. Ramani [55] | Heuristic | Substitution | √ | | √ | √ | | | 2,3 |
| 2010 | X. Huang, Y Xie [32] | BESO | Lagrangian multiplier | √ | | √ | | | | 2 |
| 2010 | X. Huang, et. al. [33] | BESO | OC | | | | | √ | | 2,3 |
| 2010 | N. Stromberg [23] | SIMP | Augmented Lagrangian | √ | | √ | | | | 2,3 |
| 2010 | S. Yamasaki, T. Nomura [56] | Level-set | Augmented Lagrangian | √ | | √ | | | | 2,3 |
| 2010 | J. Rong, J. Yi [24] | SIMP | Quadratic programming | | √ | | | | | 2 |
| 2011 | A. Gersborg [21] | SIMP | MMA | √ | | | | | | 2 |
| 2012 | M. Bruggi, P. Duysinx [19] | SIMP | MMA | | | √ | √ | | | 2 |
| 2013 | T. Liu, S. Wang [35] | Level-set | OC | √ | | | | | | 2 |
| 2013 | M. Wang, L. Li [39] | Level-set | Lagrangian multiplier | | | | √ | | | 2 |
| 2013 | G. Allaire [40] | Level-set | Augmented Lagrangian | √ | | √ | | | | 2,3 |
| 2013 | K. Suresh, et.al. [36] | Level-set | Adaptive weight | | | √ | √ | | | 2,3 |
| 2014 | S. Deng, K. Suresh [38] | Level-set | Augmented Lagrangian | | √ | √ | √ | | | 2 |
| 2014 | P. Dunning, B. Stanford [41] | Level-set | Lagrangian multiplier | | | √ | | | | 3 |
| 2014 | E. Holmberg [25] | SIMP | MMA | | | | √ | | | 2 |
| 2015 | P. Coelho, H. Rodrigues [26] | SIMP | MMA | √ | √ | | | | | 2,3 |
| 2015 | J. Li, et. al. [27] | SIMP | Multi-point Approximation | | | | | √ | | 3 |
| *Proposed* | *S. Deng, K. Suresh* | *Level-set* | *Augmented Lagrangian* | √ | √ | √ | √ | √ | √ | *3* |

To illustrate the concept of topological sensitivity, consider the 2-D example illustrated earlier in Figure 1. Assume that the quantity of interest is $Q$ (example: compliance $J$). Suppose an infinitesimal hole of radius $r$ is inserted into the domain as illustrated in Figure 3, one can expect that this will perturb the finite element solution $u$ and the quantity $Q$. The topological sensitivity of $Q$ (i.e., topological derivative) is defined in 2-D as [43]:

$$\mathcal{T}_Q(p) \equiv \lim_{r \to 0} \frac{Q(r) - Q}{\pi r^2} \quad (3.1)$$

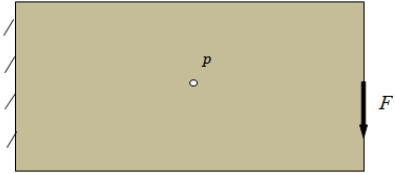

**Figure 3:** A topological change.

Starting from the definition in Equation (3.1), one can derive a closed-form expression for the topological sensitivity of compliance [57]:

$$\mathcal{T}_J(p) = \frac{4}{1+\nu} \sigma : \varepsilon - \frac{1-3\nu}{1-\nu^2} tr(\sigma) tr(\varepsilon) \quad (3.2)$$

where $\sigma$ is the stress tensor, and $\varepsilon$ is the strain tensor. Observe that the topological sensitivity is a field defined at all points within the domain. For the problem in Figure 1, the field is illustrated (after scaling) in Figure 4. Observe from the definition, and in Figure 4, that regions with relatively high values of the field correspond to regions of significant importance to the quantity of interest.

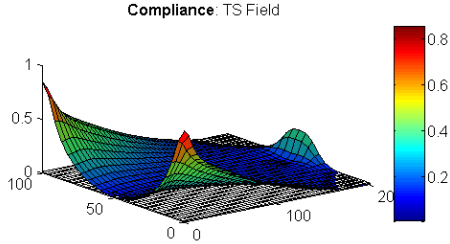

**Figure 4:** Compliance topological sensitivity.

Similar topological sensitivity expressions can be derived for other quantities of interest such as the p-norm stress [36]:

$$\mathcal{T}_\sigma(p) = \frac{4}{1+\nu} \sigma : \varepsilon - \frac{1-3\nu}{1-\nu^2} tr(\sigma) tr(\lambda) \quad (3.3)$$

where $\lambda$ is the adjoint field associated with the p-norm stress. For eigen-value problems, one can show the topological sensitivity field of eigen-value is given by [58], [59]:

$$\mathcal{T}_\lambda(p) = \sigma : \varepsilon - \omega_n^2 \rho \|u_n\|^2 \quad (3.4)$$

where $\rho$ is the material density, $\omega_n$ is the $n^{th}$ eigen-value and $u_n$ is the corresponding eigen-vector.

In instances where finding a closed-form expression for the topological sensitivity is difficult (for example, buckling), a numerical approximation may be obtained via element-sensitivity [60]. The sensitivity field of buckling load factor in linear buckling analysis can be shown to be [61]:

$$\mathcal{T}_P(p) = u^T \left( K^e - \lambda K_\sigma^e \right) u \quad (3.5)$$

where $K^e$ is element stiffness matrix, $K_\sigma^e$ is element geometric stiffness matrix, $\lambda$ is the buckling critical load and $u$ is the corresponding buckling mode shape.

### 3.2 Topological Level-Set

One approach to exploiting topological sensitivity in topology optimization is to use this field as a guide for introducing holes into an auxiliary level-set [37]. An alternate approach proposed in [62], [63] is to directly use the topological sensitivity field as a level-set; this is the approach taken in this paper.

Interpreting the field in Figure 4 as a level-set, one can introduce a cutting-plane to extract a topology. For example, using a cutting plane with an arbitrary value of $\tau = 0.03$, Figure 5 illustrates the corresponding topology $\Omega^\tau$ that is extracted via:

$$\Omega^\tau = \{ p \mid \mathcal{T}_J(p) > \tau \} \quad (3.6)$$

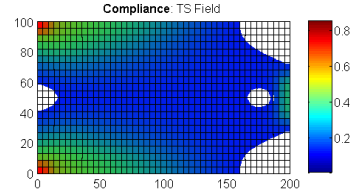

**Figure 5:** Topology extraction from the topological sensitivity field.

However, a naïve use of Equation (3.6) to carry out TO will be unstable. Instead the robust algorithm described in [62], [63] is employed by tracing the Pareto curve starting from a volume fraction of 1.0. For an incremental target volume fraction (say 0.95), a fixed point iteration method [64] is carried (see Figure 6) consisting of three steps: (1) solving the finite element problem over the current topology to obtain displacement fields; (2) calculating the topological sensitivity, and (3) extracting the new topology from topological level-set for the target volume fraction. The above three steps are repeated until convergence is reached; in practice, 2~4 iterations are sufficient to achieve convergence [62], [63]. Once convergence is reached, the target volume fraction is decremented (to say, 0.90), and the process is repeated until further volume reduction is not possible.

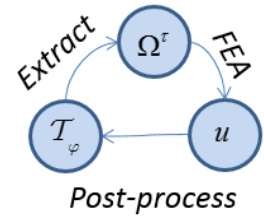

**Figure 6:** Fixed point iteration involving three quantities

### 3.3 Augmented Lagrangian Method

We now combine the above TO formulation with augmented Lagrangian method to handle constraints. Towards this end, consider the classic *continuous variable* constrained optimization problem:

$$\begin{aligned} \underset{x}{Min}\ & f(x) \\ & g_i(x) \leq 0; i = 1, 2, ..., m \end{aligned} \quad (3.7)$$

One of the most popular methods for solving such constrained optimization problems is the augmented Lagrangian method [9]. In this method, the constraints are absorbed into the objective function as follows:

$$L(x, \mu, \gamma) = f(x) + \sum_{i=1}^{m} \bar{L}_i(x, \mu, \gamma) \quad (3.8)$$

where

$$\bar{L}_i(x, \mu, \gamma) = \begin{cases} \mu_i g_i + \frac{1}{2}\gamma_i(g_i)^2; & \mu_i + \gamma_i g_i(x) > 0 \\ \frac{1}{2}\mu_i^2 / \gamma_i & \mu_i + \gamma_i g_i(x) \leq 0 \end{cases} \quad (3.9)$$

$\mu_i$ : Lagrangian multipliers
$\gamma_i$ : Penalty parameters  (3.10)

When constraints are active, the augmented Lagrangian is the combination of the linear and quadratic terms, else it takes a constant value depending on the algorithm parameters. Please see [9] for a discussion of the underlying theory.

The Lagrangian multipliers and penalty parameters are initialized to an arbitrary set of positive values. Then the augmented Lagrangian in Equation (3.8) is minimized using, for example, conjugate gradient method. In every iteration, the Lagrangian multipliers are updated as follows:

$$\mu_i^{k+1} = \max\{\mu_i^k + \gamma_i g_i(\hat{x}^k), 0\}, i = 1, 2, 3, ..., m \quad (3.11)$$

where the $\hat{x}^k$ is the local minimum at the current $k$ iteration. The penalty parameters are updated via:

$$\gamma_i^{k+1} = \begin{cases} \gamma_i^k & \min(g_i^{k+1}, 0) \leq \varsigma \min(g_i^k, 0) \\ \max(\eta \gamma_i^k, k^2) & \min(g_i^{k+1}, 0) > \varsigma \min(g_i^k, 0) \end{cases} \quad (3.12)$$

where $0 < \varsigma < 1$ and $\eta > 0$; typically $\varsigma = 0.25$ and $\eta = 10$ [9].

Upon updating, the augmented Lagrangian is once again minimized, and the process is repeated until termination.

### 3.4 Augmented Topological Level-Set

The goal now is to extend the above augmented Lagrangian method to solve the constrained TO problem in Equation (1.1). Drawing an analogy between Equations (3.7) and (1.1), we define the *topological augmented Lagrangian* as follows:

$$L(u, \Omega; \gamma_i, \mu_i) \equiv \varphi + \sum_{i=1}^{m} \bar{L}_i(u, \Omega; \gamma_i, \mu_i) \quad (3.13)$$

where

$$\bar{L}_i(u, \Omega; \gamma_i, \mu_i) = \begin{cases} \mu_i g_i + \frac{1}{2}\gamma_i(g_i)^2 & \mu_i + \gamma_i g_i > 0 \\ \frac{1}{2}\mu_i^2 / \gamma_i & \mu_i + \gamma_i g_i \leq 0 \end{cases} \quad (3.14)$$

In classic continuous optimization, the gradient of the augmented Lagrangian in Equation (3.8), with respect to the continuous variable $x$, is given by:

$$\nabla L(x, \mu, \gamma) = \nabla f + \sum_{i=1}^{m} \nabla \bar{L}_i(x, \mu, \gamma) \quad (3.15)$$

where

$$\nabla \bar{L}_i(x, \mu, \gamma) = \begin{cases} \mu_i + \gamma_i g_i\ \nabla g_i & \mu_i + \gamma_i g_i(x) > 0 \\ 0 & \mu_i + \gamma_i g_i(x) \leq 0 \end{cases} \quad (3.16)$$

Here, the gradient is defined with respect to a topological change. Drawing an analogy to the gradient operator in Equation (3.15), the *topological gradient operator* is defined as

$$\mathcal{T}_\Omega[L(u, \Omega; \gamma_i, \mu_i)] \equiv \mathcal{T}_L = \mathcal{T}_\varphi + \sum_{i=1}^{m} \mathcal{T}_{\bar{L}_i} \quad (3.17)$$

where $\mathcal{T}_\varphi$ is the topological level-set associated with the objective, and

$$\mathcal{T}_{\bar{L}_i} = \begin{cases} \mu_i + \gamma_i g_i\ \mathcal{T}_{g_i} & \mu_i + \gamma_i g_i > 0 \\ 0 & \mu_i + \gamma_i g_i \leq 0 \end{cases} \quad (3.18)$$

where

$$\mathcal{T}_{g_i} \equiv \mathcal{T}(g_i) \quad (3.19)$$

are the topological level-sets associated with each of the constraint functions. Observe that we have essentially combined various topological sensitivities into a single topological level set. The multipliers and penalty parameters are updated as described earlier, and the complete algorithm is described in a later section.

### 3.5 Assembly-Free Deflated Conjugate Gradient Method

A practical challenge that arises in 3D constrained TO is the computational cost. Specifically, for each of the constraints one must solve a distinct finite element problem, and compute the corresponding topological sensitivity.

To address the computational cost, we rely here on assembly-free methods. Specifically, for solving the static finite element problem, we use the assembly-free deflated conjugate gradient (AF-DCG) method proposed in [42]. The AF-DCG rests on the observation that the computational bottle-neck in modern architecture is memory access [65]. AF-DCG computes the preconditioner and the solution to the underlying linear system in an assembly-free manner, significantly reducing memory bandwidth. Similarly, for modal analysis, an assembly-free modal analysis proposed in [66] is employed. Finally, for linear buckling analysis, the assembly-free extension of this method to buckling [67] is used.

While the above papers address each of the finite element problems individually, in this paper, they are effectively combined to solve constrained TO problems.

### 3.6 Hard and Soft Constraints

In classic optimization [9], constraints are typically treated as 'hard' constraints, i.e., the algorithm will terminate if any of the constraints is violated. In design optimization, this can be too restrictive since design constraints may be unreasonable, and the algorithm may terminate without any solution.

Researchers have therefore developed algorithms for handling of soft constraints [68]: "*Hard constraints limit the feasible space, while soft constraints prioritize solutions within this space.*" Soft constraints are particularly useful in engineering [69] as an alternate to multi-objective problems.

In this study, we permit hard and soft constraints; for example:

$$\underset{\Omega \subset D}{Min} \varphi$$
$$g_1(u,\Omega) \leq 0 \quad (3.20)$$
$$g_2(u,\Omega) \leq 0 \quad \text{(soft)}$$

The first constraint is hard, while the second constraint is treated as soft. Soft constraints influence the topology through the Lagrangian multiplier (see algorithm below), but do not influence the termination of the algorithm, i.e., do not influence the feasible space. Typically compliance and stress are treated as hard constraints, while buckling and Eigen-value can be treated as either hard or soft constraints.

**3.7 Casting Constraints**

In addition to performance constraints, it is often important to include manufacturing constraints. For example, in [40] the authors proposed a projection method within a level-set formulation to impose thickness constraint for cast parts. Similarly, in [70], the authors imposed a density constraint, within the SIMP formulation, along the casting direction. This ensures that the density variable is non-decreasing along the casting direction (on either sides of the parting plane), preventing cavities in cast parts. In this paper, we adopt this method to the topological level-set formulation; specifically, after a casting direction is selected, we impose a constraint on topological level-set to be non-decreasing along the casting direction to prevent cavities.

**3.8 Proposed Algorithm**

The four performance constraints considered here are compliance ($J$), p-norm Von Mises stress ($\sigma$), lowest Eigen-value ($\lambda$) and buckling load ($P$); soft constraints are identified, and casting constraints are optional.

The constraints are typically set relative to their initial values prior to optimization, and the objective is to minimize volume, i.e., the generic problem considered here is:

$$\underset{\Omega \subset D}{Min} |\Omega|$$
$$J \leq (\alpha_1) J_0$$
$$\sigma \leq (\alpha_2) \sigma_0$$
$$P \geq (\alpha_3) P_0 \quad (\alpha_3 < 1)$$
$$\lambda \geq (\alpha_4) \lambda_0 \quad (\alpha_4 < 1) \quad (3.21)$$
$$P \geq (\alpha_5) P_0 \quad \text{(soft)}$$
$$\lambda \geq (\alpha_6) \lambda_0 \quad \text{(soft)}$$
$$Ku = f$$

Thus, the constraint in Equation (3.22) implies that the final compliance must not exceed three times the initial compliance.

$$J \leq 3J_0 \quad (3.22)$$

Similarly, Equation (3.23) implies that the final p-norm von Mises stress must not exceed twice the initial p-norm stress.

$$\sigma \leq 2\sigma_0 \quad (3.23)$$

On the other hand, the buckling and Eigen-value constraints may be hard or soft. For a soft constraint

$$P \geq 1.2P_0 \quad \text{(soft)} \quad (3.24)$$

the algorithm will attempt to find solutions (within the feasible space) that satisfy the above equation. Observe that, if one imposes a hard constraint:

$$P \geq 1.2P_0 \quad (3.25)$$

the algorithm will terminate at the first iteration since the initial design will not satisfy this constraint!

The overall algorithm is illustrated in Figure 7, and it proceeds as follows:

1. The domain is discretized using hexahedral elements; the optimization parameters are initialized as $\mu = 100$ and $\gamma = 10$.
2. Depending on the constraint imposed, several FEAs are performed.
3. The constraints are evaluated, and the Lagrangian parameters are updated.
4. If any of the *hard* constraints are violated, the algorithm proceeds to step-9, else, it proceeds to step-5.
5. The topological sensitivity fields are computed, and the augmented topological level-set is extracted.
6. The topology for the current volume fraction is extracted.
7. If the topology has converged (i.e., if the change in compliance is less than 1%), proceed to step-8, else return to step-2.
8. Decrement the target-volume fraction $v = v - \Delta v$, and return to step-2.
9. Decrease the volume step-size $\Delta v$; if the step-size is smaller than $\Delta v_{\min}$, terminate the algorithm, else return to step-2.

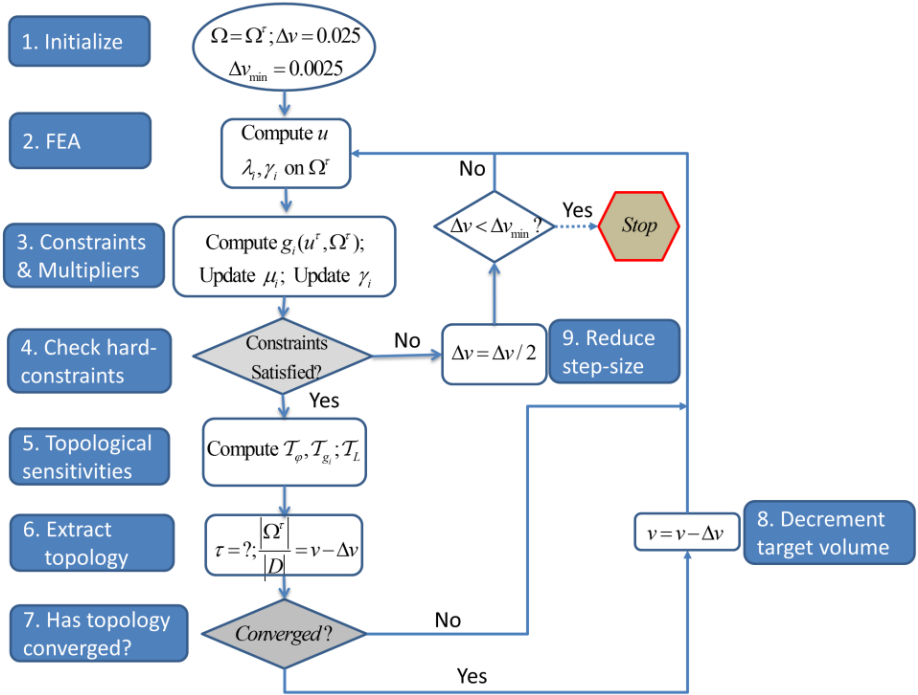

**Figure 7:** Proposed algorithm.

## 4. NUMERICAL EXAMPLES

In this section, numerical experiments are carried out to illustrate the above algorithm; the default parameters are:

- Material properties: $E = 2*10^{11}$ and $\nu = 0.3$
- All experiments were conducted using C++ on a Windows 7 64-bit machine with the following hardware: Intel I7 960 CPU quad-core running at 3.2GHz with 6 GB of memory.

### 4.1 L-bracket with Tip Load

The first experiment involves the L-bracket, whose cross-section is illustrated in Figure 8 (units in mm), with a thickness of 6 mm. In the TO literature, it is common to use an L-bracket with a sharp reentrant corner [71]. This is perfectly acceptable for compliance dominated problems, but may not be desirable for stress-constrained problems due to the stress singularity at the reentrant corner. We have therefore added a small fillet to relieve the stress singularity. The L-bracket is fixed on the top edge, while a unit load is applied as illustrated. The domain is discretized with 24,330 elements, i.e., 90,738 degrees of freedom (DOF).

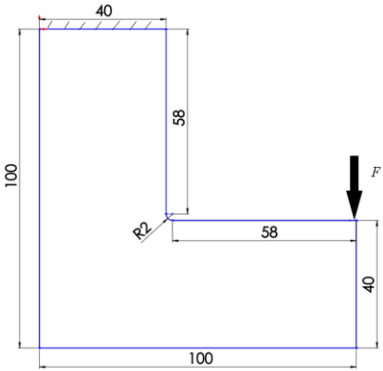

**Figure 8:** L-bracket model.

The specific constrained TO problem considered here is:

$$\underset{\Omega \subset D}{Min} |\Omega|$$
$$J \leq (\alpha_1) J_0 \qquad (4.1)$$
$$\sigma \leq (\alpha_2) \sigma_0$$

Two scenarios are summarized in Table 2; the first scenario is compliance dominant, while the second is stress dominant. The final results, volume fractions and computing time are also summarized in Table 2. The active constraints are identified with a 'box'.

**Table 2:** Constraints and results for problem in Figure 8.

| Dominant Constraint | Initial Constraints | Final Results | Final volume & running time (s) |
|---|---|---|---|
| Compliance | $J \leq 2J_0$ | $\boxed{J = 2J_0}$ | $v = 0.34$ |
|  | $\sigma \leq 100\sigma_0$ | $\sigma = 1.27\sigma_0$ | $t = 142.11$ |
| Stress | $J \leq 100J_0$ | $J = 1.75J_0$ | $v = 0.47$ |
|  | $\sigma \leq 1.05\sigma_0$ | $\boxed{\sigma = 1.05\sigma_0}$ | $t = 194.32$ |

The corresponding optimized topologies are illustrated in Figure 9. Observe that when compliance is dominant, the classic stiff design is obtained, whereas when stress is dominant, a strong design is obtained where the fillet radius is increased to reduce stress [72].

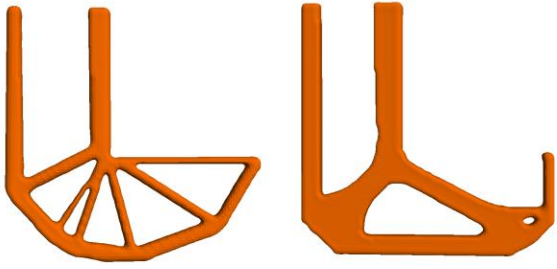

**Figure 9:** Final topologies which are subject to dominant constraints of compliance (left) and Von Mises stress (right).

Figure 10 illustrates the relative cost of various sections of the algorithm. As one can observe, significant portion (88%) of the computational time is spent on FEA, while the remaining 12% is spent on computing the topological sensitivity field, and other tasks.

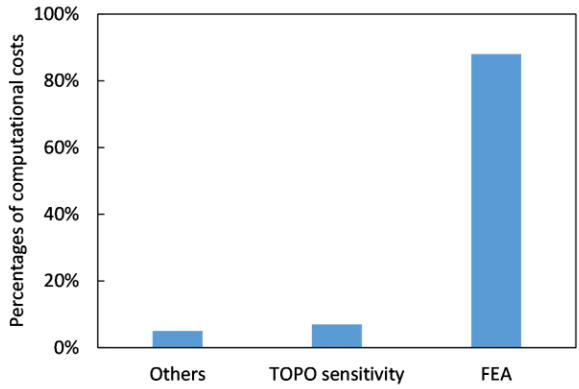

**Figure 10:** Computational cost for scenario-1 in Table 2.

### 4.2 Plate with Pressure Load

In the next example, we consider a plate geometry whose cross-section is illustrated in Figure 11 (units in mm); the thickness is 10 mm. The left face is fixed while a unit horizontal pressure is applied on the right face. The geometry is meshed with 50,560 hexahedral elements, i.e., 167,280 DOF.

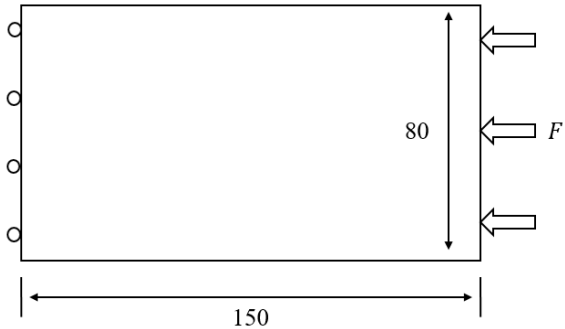

**Figure 11:** Thick plate dimensions and pressure loading.

The specific constrained TO problem considered here is:

$$\underset{\Omega \subset D}{Min} |\Omega|$$
$$J \leq (\alpha_1) J_0$$
$$\sigma \leq (\alpha_2) \sigma_0 \qquad (4.2)$$
$$P \geq (\alpha_3) P_0$$

Three instances are summarized in Table 3; once again, the active final constraints are identified with a box. The execution time for the buckling-dominated problem is much longer due to the inherent computational complexity.

**Table 3:** Constraints and results for problem in Figure 11.

| Dominant Constraint | Initial Constraints | Final Results | Final volume & running time (s) |
|---|---|---|---|
| Compliance | $J \leq 5J_0$ | $\boxed{J = 5J_0}$ | $v = 0.22$ |
| | $\sigma \leq 100\sigma_0$ | $\sigma = 2.39\sigma_0$ | $t = 342.63$ |
| | $P \geq 0.1P_0$ | $P = 0.16P_0$ | |
| Stress | $J \leq 100J_0$ | $J = 5.25J_0$ | $v = 0.22$ |
| | $\sigma \leq 2\sigma_0$ | $\boxed{\sigma = 2\sigma_0}$ | $t = 401.09$ |
| | $P \geq 0.1P_0$ | $P = 0.11P_0$ | |
| Buckling | $J \leq 100J_0$ | $J = 1.90J_0$ | $v = 0.71$ |
| | $\sigma \leq 100\sigma_0$ | $\sigma = 1.73\sigma_0$ | $t = 234.34$ |
| | $P \geq 0.9P_0$ | $\boxed{P = 0.9P_0}$ | |

The corresponding topologies are illustrated in Figure 12. As one can observe, the topologies for the first two cases are similar, and this is consistent with the results in Table 3. The topology for the buckling-dominated problem is, however, significantly different.

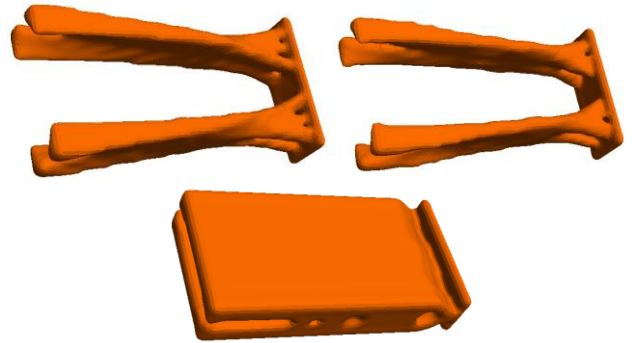

**Figure 12:** Final topologies for compliance dominated (top) stress dominated (left) and buckling-dominated (right).

In order to study the termination criterion, the iteration history for scenario-2 is illustrated in Figure 13 (the plot is to be interpreted from the right to left). Observe that the optimization starts with a volume decrement of 0.025, and it reduces when divergence is detected; the decrement increases slightly towards the end. Most importantly, the volume decrement is well above the minimum value during optimization. Thus, the termination is triggered by the hard stress constraint rather than the volume decrement constraint.

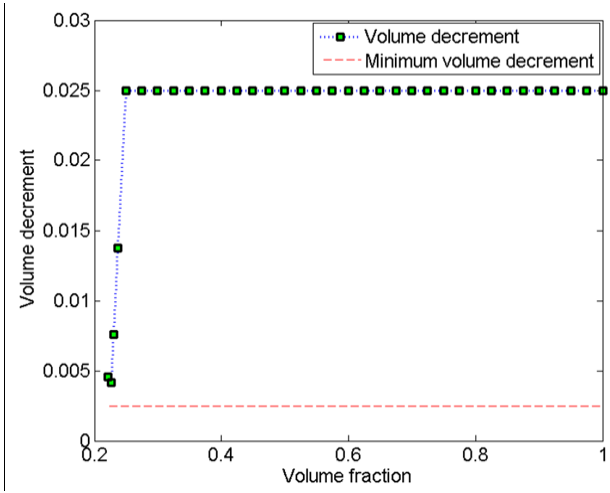

**Figure 13:** Iteration history of volume decrement for the scenario-2 in Table 3.

For scenario-2 in Table 3, the relative constraints are illustrated Figure 14 (the plots are to be interpreted from the right to left). The optimization terminates due to the stress constraint at a final volume fraction of 0.22.

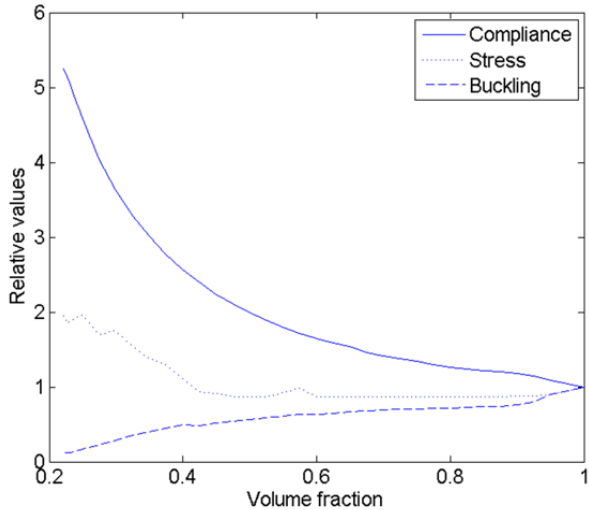

**Figure 14:** Constraint iteration history for the scenario-2 in the Table 3.

### 4.3 Case Study: Flange

This case study involves the flange illustrated in Figure 15; units are in inches. Flanges are commonly used, for example, to fasten pipes and rail-joints. The objective is to minimize the flange weight while subject to compliance, stress and Eigen-mode constraint. For FEA, 19,924 hexahedral elements are used to discretize the design domain, resulting in 63,666 DOF.

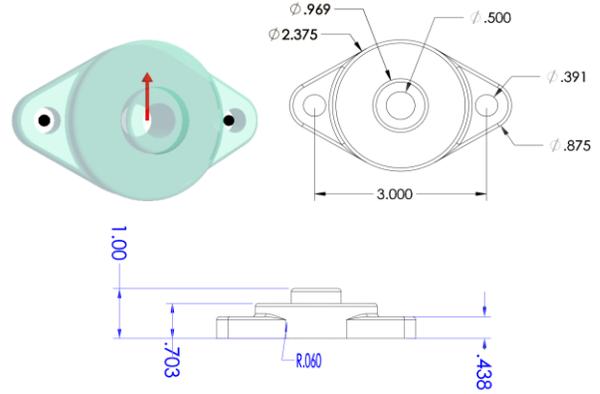

**Figure 15:** Flange structure and dimensions.

The specific constrained TO problem considered here is:

$$\underset{\Omega \subset D}{Min} |\Omega|$$
$$J \leq (\alpha_1) J_0$$
$$\sigma \leq (\alpha_2) \sigma_0 \qquad (4.3)$$
$$\lambda \geq (\alpha_4) \lambda_0 \quad \text{(soft, if } \alpha_4 > 1\text{)}$$

The Eigen-mode constraint is soft if the corresponding multiplier is greater than 1, to avoid early termination.

Table-4 summarizes the results for 3 different scenarios. In particular, for scenario-3, observe that although the Eigen-value constraint is soft, its impact on the final result and topology is self-evident.

**Table 4:** Constraints and results for problem in Figure 15.

| Dominant Constraint | Initial Constraints | Final Results | Final volume & running time (s) |
|---|---|---|---|
| Compliance | $J \leq 2J_0$<br>$\sigma \leq 100\sigma_0$<br>$\lambda \geq 0.1\lambda_0$ | $\boxed{J = 2J_0}$<br>$\sigma = 1.12\sigma_0$<br>$\lambda = 0.69\lambda_0$ | $v = 0.44$<br>$t = 156.91$ |
| Stress | $J \leq 100J_0$<br>$\sigma \leq 1.05\sigma_0$<br>$\lambda \geq 0.1\lambda_0$ | $J = 4.66J_0$<br>$\boxed{\sigma = 1.05\sigma_0}$<br>$\lambda = 1.27\lambda_0$ | $v = 0.44$<br>$t = 185.97$ |
| Stress and Eigen-value | $J \leq 100J_0$<br>$\sigma \leq 1.2\sigma_0$<br>$\lambda \geq 1.5\lambda_0$ (soft) | $J = 1.84J_0$<br>$\boxed{\sigma = 1.20\sigma_0}$<br>$\boxed{\lambda = 2.49\lambda_0}$ | $v = 0.68$<br>$t = 77.22$ |

The corresponding topologies are illustrated in Figure 16. The volume fractions for the first two scenarios are identical, but the difference in topology is worth noticing (also see Table-4).

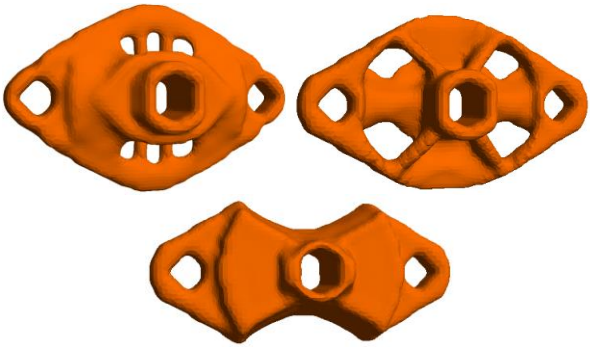

**Figure 16:** Final topologies for compliance dominated (top-left) stress dominated (top-right), and stress and Eigen-value dominated (bottom).

In order to study the iteration history of the soft-constrained optimization, the volume decrement history is plotted in Figure 17 for scenario-3 in Table 4. It is seen the volume decrement was kept constant until the last two steps when convergence error was detected.

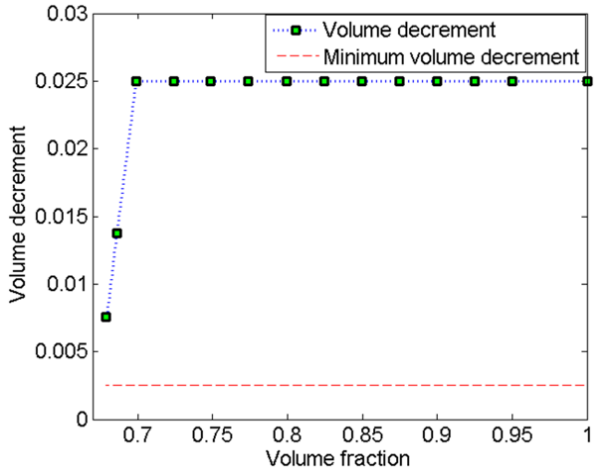

**Figure 17:** Volume decrement history for the scenario-3 in the Table 4.

The constraint iteration history for scenario-3 in Table 4 is illustrated in Figure 18 where the lowest eigen-value sees a significant increase with material removal. However, since the eigen-value constraint is 'soft', it does not lead to early termination. The optimization terminated due to a hard stress constraint of 1.2 at a final volume of 0.68.

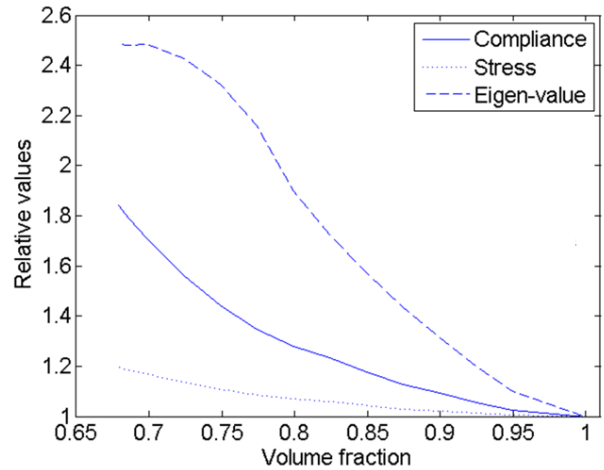

**Figure 18:** Iteration history of constraints for the scenario-3 in Table 4.

### 4.4 Case Study: Bicycle Frame

In this case study, we use the proposed algorithm to find a conceptual design for a bicycle frame. The design space is illustrated in Figure 19 where all units are in centimeters.

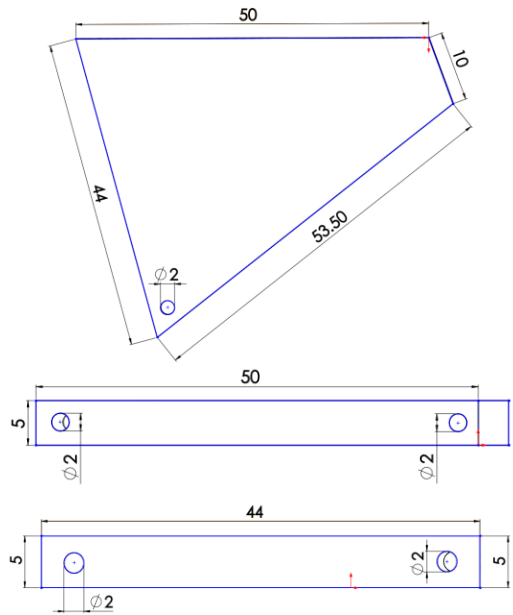

**Figure 19:** Design space of bike frame: front view (1st), top view (2nd) and side view (3rd).

Two loads are applied as in Figure 20, where $F_1$ is 60 N, and $F_2$, is 140 N; see [73]. The two loads act simultaneously, i.e., this is a *multiple-load problem*. The design is discretized into 51,280 hexahedral elements, i.e., 176,367 DOFs.

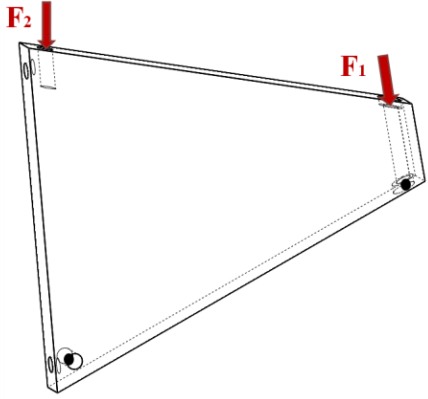

**Figure 20:** The bike frame subject to multiple loads.

Only one scenario is considered; the constraints and final results are summarized in Table 5.

**Table 5:** Constraints and results for problem in Figure 20.

| Dominant Constraint | Initial Constraints | Final Results | Final vol. & time (s) |
|---|---|---|---|
| Compliance, Stress and Buckling | $J \leq 20 J_0$ <br> $\sigma \leq 20\sigma_0$ <br> $P \geq 1.5 P_0 (soft)$ | $\boxed{J = 19.47 J_0}$ <br> $\sigma = 11.92\sigma_0$ <br> $P = 0.08 P_0$ | $v = 0.22$ <br> $t = 953.58$ |

The final design is illustrated in Figure 21.

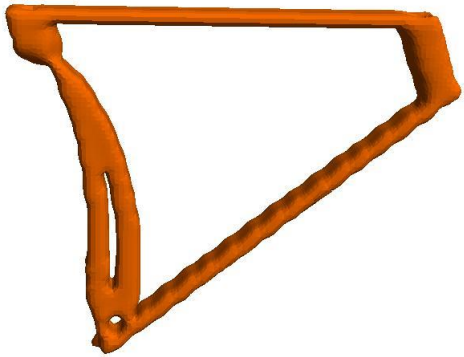

**Figure 21:** Proposed design for a bike frame.

### 4.5 Case Study: Bicycle crank

In the final case-study, we optimize the design of a bicycle crank arm. The 2-D sketch of the design space is illustrated in Figure 22 (units in mm), with a thickness of 15 mm. The structure is discretized using 36,608 elements, with 128,250 DOFs.

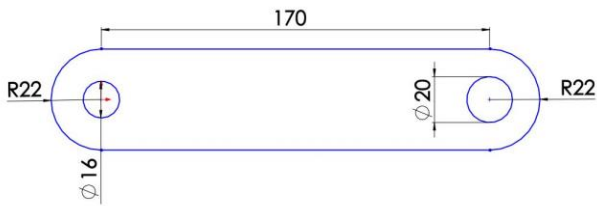

**Figure 22**: Dimensions of the crank arm.

In this example, a *multi-load* scenario is considered, i.e., during a full pedaling cycle, the crank arm passes through four distinct positions as illustrated in Figure 23. At each position, it experiences a different loading condition. At position A where the pedal is passing through the highest point, it sees a compressive load. At position B where the crank arm is horizontally placed, it sees a bending load. At position C, it sees a tension force. At position D, the load is negligible. The magnitudes of pedaling forces $F_1$, $F_2$ and $F_3$ are in the ratio 1:5:2.2, with $F_1$ being 50 N [74].

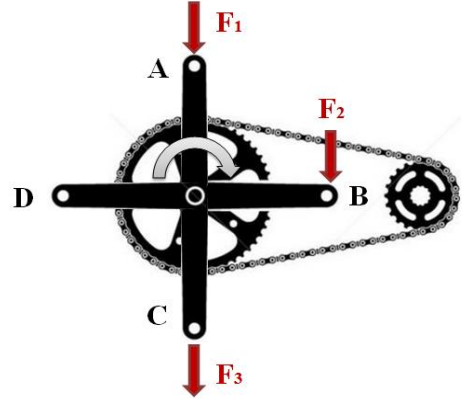

**Figure 23:** The crank arm subject to multi-load during a pedaling cycle [74].

The objective is to minimize the weight subject to the constraints summarized in Table 6; stiffness and strength being the most important constraints. A soft buckling constraint is imposed for the compressive load $F_1$. Thus, one can impose different constraints for different sets of loads.

The optimization results are summarized in Table 6 where it is noted that the compliance constraints for loads $F_1$ and $F_3$ are active at termination. Since this is a multi-load problem, the computational cost if fairly high (about 25 minutes), despite the use of fast assembly-free methods.

**Table 6:** Constraints and results for problem in Figure 22.

| Loads | Initial Constraints | Final Results | Final vol. & time (s) |
|---|---|---|---|
| Compression $F_1$ | $J \leq 1.5 J_0$ <br> $\sigma \leq 4\sigma_0$ <br> $P \geq 5 P_0 (soft)$ | $\boxed{J = 1.50 J_0}$ <br> $\sigma = 1.33\sigma_0$ <br> $P = 0.75 P_0$ | $v = 0.69$ <br> $t = 1507.29$ |
| Bending $F_2$ | $J \leq 1.5 J_0$ <br> $\sigma \leq 4\sigma_0$ | $J = 1.25 J_0$ <br> $\sigma = 1.01\sigma_0$ | |
| Tension $F_3$ | $J \leq 1.5 J_0$ <br> $\sigma \leq 4\sigma_0$ | $\boxed{J = 1.50 J_0}$ <br> $\sigma = 1.33\sigma_0$ | |

The final topology is illustrated in Figure 24. Although the design meets the performance constraints, it exhibits 'undercuts', i.e., cavities. This may not be desirable if the part needs to be cast.

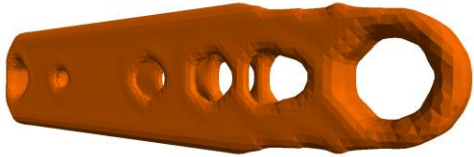

**Figure 24:** Final design of crank arm.

We therefore imposed a casting constraint (through the thickness) in addition to the performance constraints; the results are summarized in Table 7. Observe that the design now not only meets the performance constraint, but also the manufacturing constraint. In this example, the impact of the manufacturing constraint on performance, and computational time, was found to be negligible.

**Table 7:** Constraints and results for problem in Figure 22.

| Loads | Initial Constraints | Final Constraints | Final vol. & time (s) |
|---|---|---|---|
| Compression $F_1$ | $J \leq 1.5 J_0$ | $\boxed{J = 1.50 J_0}$ | $v = 0.70$ |
|  | $\sigma \leq 4\sigma_0$ | $\sigma = 1.30\sigma_0$ | $t = 1489.59$ |
|  | $P \geq 5 P_0 (soft)$ | $P = 0.72 P_0$ |  |
| Bending $F_2$ | $J \leq 1.5 J_0$ | $J = 1.23 J_0$ |  |
|  | $\sigma \leq 4\sigma_0$ | $\sigma = 1.01\sigma_0$ |  |
| Tension $F_3$ | $J \leq 1.5 J_0$ | $\boxed{J = 1.50 J_0}$ |  |
|  | $\sigma \leq 4\sigma_0$ | $\sigma = 1.30\sigma_0$ |  |

The resulting design is illustrated in Figure 25; observe that the design does not exhibit undercuts.

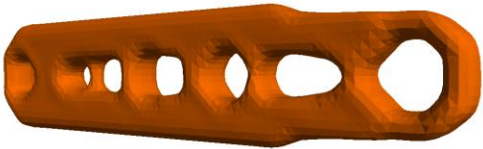

**Figure 25:** Final design of crank arm with casting constraint.

## 5. CONCLUSIONS

The proposed method inherits the robustness and generality of the classic augmented Lagrangian method. Specifically, through several numerical experiments, we demonstrated that the proposed method can solve a variety of multi-constrained (single-load, multiple-load and multi-load) TO problems. By varying the constraint limits, we were able to explore the impact of these constraints on the final topology. We were also able to explore the impact of manufacturing constraint on the topology. Future work will explore the generalization of this method to multi-physics problems.

**Acknowledgements**

The authors would like to thank the support of National Science Foundation through grants CMMI-1232508, CMMI-1161474, and IIP-1500205.

**References**

[1] M. Bendsøe and O. Sigmund, *Topology Optimization: Theory, Methods and Application*, 2nd ed. Springer, 2003.
[2] S. Nishiwaki, "Topology Optimization of Compliant Mechanisms using the Homogenization Method," *International Journal for Numerical Methods in Engineering*, vol. 42, pp. 535–559, 1998.
[3] O. Sigmund, "A 99 line topology optimization code written in Matlab," *Structural and Multidisciplinary Optimization*, vol. 21, no. 2, pp. 120–127, 2001.
[4] Y. L. Mei, "A level set method for structural topology optimization and its applications," *Advances in Engineering Software*, vol. 35, no. 7, pp. 415–441, 2004.
[5] Q. Xia, T. Shi, S. Liu, and M. Y. Wang, "A level set solution to the stress-based structural shape and topology optimization," *Computers & Structures*, vol. 90–91, pp. 55–64, 2012.
[6] K. Svanberg, "The method of moving asymptotes—a new method for structural optimization," *Int. J. Numer. Meth. Engng.*, vol. 24, no. 2, pp. 359–373, Feb. 1987.
[7] J. Sokolowski, "Optimality Conditions for Simultaneous Topology and Shape Optimization," *SIAM Journal on Control and Optimization*, vol. 42, no. 4, pp. 1198–1221, 2003.
[8] J. Arora, *Introduction to optimum design*. New York: Academic Press, 2004.
[9] J. Nocedal and S. Wright, *Numerical Optimization*. Springer, 1999.
[10] M. Kočvara, "Topology optimization with displacement constraints: a bilevel programming approach," *Structural Optimization*, vol. 14, no. 4, pp. 256–263, Dec. 1997.
[11] K. Mela, "Resolving issues with member buckling in truss topology optimization using a mixed variable approach," *Struct Multidisc Optim*, vol. 50, no. 6, pp. 1037–1049, Jul. 2014.
[12] C. Ni, J. Yan, G. Cheng, and X. Guo, "Integrated size and topology optimization of skeletal structures with exact frequency constraints," *Struct Multidisc Optim*, vol. 50, no. 1, pp. 113–128, Jan. 2014.
[13] G. I. N. Rozvany, "A critical review of established methods of structural topology optimization," *Structural and Multidisciplinary Optimization*, vol. 37, no. 3, pp. 217–237, 2009.
[14] M. P. Bendsøe, "Optimal shape design as a material distribution problem," *Structural Optimization*, vol. 1, no. 193–202, 1989.
[15] R. J. Yang and C. J. Chen, "Stress-Based Topology Optimization," *Structural Optimization*, vol. 12, pp. 98–105, 1996.
[16] V. K. Yalamanchili and A. V. Kumar, "Topology Optimization of Structures using a Global Stress Measure," in *Proceedings of the ASME 2012 International Design Engineering Technical Conferences & Computers and Information in Engineering Conference*, Chicago, IL, 2012.
[17] M. Werme, "Using the sequential linear integer programming method as a post-processor for stress-constrained topology optimization problems," *Int. J. Numer. Meth. Engng.*, vol. 76, no. 10, pp. 1544–1567, Dec. 2008.

[18] M. Bruggi and P. Venini, "A mixed FEM approach to stress-constrained topology optimization," *Int. J. Numer. Meth. Engng.*, vol. 73, no. 12, pp. 1693–1714, Mar. 2008.

[19] M. Bruggi and P. Duysinx, "Topology optimization for minimum weight with compliance and stress constraints," *Structural and Multidisciplinary Optimization*, vol. 46, pp. 369–384, 2012.

[20] J. T. Pereira, E. A. Fancello, and C. S. Barcellos, "Topology optimization of continuum structures with material failure constraints," *Struct Multidisc Optim*, vol. 26, no. 1–2, pp. 50–66, Sep. 2003.

[21] A. R. Gersborg and C. S. Andreasen, "An explicit parameterization for casting constraints in gradient driven topology optimization," *Struct Multidisc Optim*, vol. 44, no. 6, pp. 875–881, Mar. 2011.

[22] K.-T. Zuo, "Manufacturing- and machining-based topology optimization," *The International Journal of Advanced Manufacturing Technology*, vol. 27, no. 5–6, pp. 531–536, 2006.

[23] N. Strömberg, "Topology optimization of structures with manufacturing and unilateral contact constraints by minimizing an adjustable compliance–volume product," *Structural and Multidisciplinary Optimization*, vol. 42, no. 3, pp. 341–350, 2010.

[24] J. H. Rong and J. H. Yi, "A structural topological optimization method for multi-displacement constraints and any initial topology configuration," *Acta Mech Sin*, vol. 26, no. 5, pp. 735–744, Aug. 2010.

[25] E. Holmberg, B. Torstenfelt, and A. Klarbring, "Fatigue constrained topology optimization," *Struct Multidisc Optim*, vol. 50, no. 2, pp. 207–219, Feb. 2014.

[26] P. G. Coelho and H. C. Rodrigues, "Hierarchical topology optimization addressing material design constraints and application to sandwich-type structures," *Struct Multidisc Optim*, vol. 52, no. 1, pp. 91–104, Jan. 2015.

[27] J. Li, S. Chen, and H. Huang, "Topology optimization of continuum structure with dynamic constraints using mode identification," *J Mech Sci Technol*, vol. 29, no. 4, pp. 1407–1412, Apr. 2015.

[28] D. J. Munk, G. A. Vio, and G. P. Steven, "Topology and shape optimization methods using evolutionary algorithms: a review," *Struct Multidisc Optim*, pp. 1–19, May 2015.

[29] X. Huang and Y. M. Xie, "A new look at ESO and BESO optimization methods," *Structural and Multidisciplinary Optimization*, vol. 35, no. 1, pp. 89–92, 2008.

[30] H. Guan, Y.-J. Chen, Y.-C. Loo, Y.-M. Xie, and G. P. Steven, "Bridge topology optimisation with stress, displacement and frequency constraints," *Computers & Structures*, vol. 81, no. 3, pp. 131–145, Feb. 2003.

[31] W. Li, Q. Li, G. P. Steven, and Y. M. Xie, "An evolutionary shape optimization procedure for contact problems in mechanical designs," *Proceedings of the Institution of Mechanical Engineers, Part C: Journal of Mechanical Engineering Science*, vol. 217, no. 4, pp. 435–446, Apr. 2003.

[32] X. Huang and Y. M. Xie, "Evolutionary topology optimization of continuum structures with an additional displacement constraint," *Struct Multidisc Optim*, vol. 40, no. 1–6, pp. 409–416, Apr. 2009.

[33] X. Huang, Z. H. Zuo, and Y. M. Xie, "Evolutionary topological optimization of vibrating continuum structures for natural frequencies," *Comput. Struct.*, vol. 88, no. 5–6, pp. 357–364, 2010.

[34] N. P. van Dijk, K. Maute, M. Langelaar, and F. van Keulen, "Level-set methods for structural topology optimization: a review," *Structural and Multidisciplinary Optimization*, vol. 48, no. 3, pp. 437–472, 2013.

[35] T. Liu, S. Wang, B. Li, and L. Gao, "A level-set-based topology and shape optimization method for continuum structure under geometric constraints," *Struct Multidisc Optim*, vol. 50, no. 2, pp. 253–273, Mar. 2014.

[36] K. Suresh and M. Takalloozadeh, "Stress-Constrained Topology Optimization: A Topological Level-Set Approach," *Structural and Multidisciplinary Optimization*, vol. 48, no. 2, pp. 295–309, 2013.

[37] G. Allaire, F. Jouve, and A. M. Toader, "Structural Optimization using Sensitivity Analysis and a Level-set Method," *Journal of Computational Physics*, vol. 194, no. 1, pp. 363–393, 2004.

[38] S. Deng and K. Suresh, "Multi-constrained topology optimization via the topological sensitivity," *Structural and Multidisciplinary Optimization*, vol. 51, no. 5, pp. 987–1001, 2015.

[39] M. Y. Wang and L. Li, "Shape equilibrium constraint: a strategy for stress-constrained structural topology optimization," *Struct Multidisc Optim*, vol. 47, no. 3, pp. 335–352, Sep. 2012.

[40] G. Allaire, F. Jouve, and G. Michailidis, "Casting constraints in structural optimization via a level-set method," in *10th World Congress on Structural and Multidisciplinary Optimization*, Orlando, United States, 2013.

[41] P. D. Dunning, B. K. Stanford, and H. A. Kim, "Coupled aerostructural topology optimization using a level set method for 3D aircraft wings," *Struct Multidisc Optim*, vol. 51, no. 5, pp. 1113–1132, Nov. 2014.

[42] P. Yadav and K. Suresh, "Large Scale Finite Element Analysis Via Assembly-Free Deflated Conjugate Gradient," *J. Comput. Inf. Sci. Eng*, vol. 14, no. 4, pp. 041008–1: 041008–9, 2014.

[43] H. A. Eschenauer, V. V. Kobelev, and A. Schumacher, "Bubble method for topology and shape optimization of structures," *Structural Optimization*, vol. 8, pp. 42–51, 1994.

[44] A. A. Novotny, R. A. Feijóo, C. Padra, and E. Taroco, "Topological Derivative for Linear Elastic Plate Bending Problems," *Control and Cybernetics*, vol. 34, no. 1, pp. 339–361, 2005.

[45] A. A. Novotny, R. A. Feijoo, and E. Taroco, "Topological Sensitivity Analysis for Three-dimensional Linear Elasticity Problem," *Computer Methods in Applied Mechanics and Engineering*, vol. 196, no. 41–44, pp. 4354–4364, 2007.

[46] I. Turevsky, S. H. Gopalakrishnan, and K. Suresh, "An Efficient Numerical Method for Computing the Topological Sensitivity of Arbitrary Shaped Features in Plate Bending," *International Journal of Numerical Methods in Engineering*, vol. 79, pp. 1683–1702, 2009.

[47] I. Turevsky and K. Suresh, "Generalization of Topological Sensitivity and its Application to Defeaturing," in *ASME IDETC Conference*, Las Vegas, 2007.

[48] M. Zhou and G. I. N. Rozvany, "DCOC: An optimality criteria method for large systems Part I: theory," *Structural Optimization*, vol. 5, no. 1–2, pp. 12–25, Mar. 1992.

[49] R. B. Haber, C. S. Jog, and M. P. Bendsøe, "A new approach to variable-topology shape design using a

constraint on perimeter," *Structural Optimization*, vol. 11, no. 1–2, pp. 1–12, Feb. 1996.
[50] J. Petersson and O. Sigmund, "Slope constrained topology optimization," *Int. J. Numer. Meth. Engng.*, vol. 41, no. 8, pp. 1417–1434, Apr. 1998.
[51] L. Yin and W. Yang, "Optimality criteria method for topology optimization under multiple constraints," *Computers & Structures*, vol. 79, no. 20–21, pp. 1839–1850, Aug. 2001.
[52] G. Allaire, F. Jouve, and A.-M. Toader, "Structural Optimization by the Level-Set Method," in *Free Boundary Problems*, P. Colli, C. Verdi, and A. Visintin, Eds. Birkhäuser Basel, 2003, pp. 1–15.
[53] M. Stolpe and T. Stidsen, "A hierarchical method for discrete structural topology design problems with local stress and displacement constraints," *International Journal for Numerical Methods in Engineering*, vol. 69, no. 5, pp. 1060–1084, Jan. 2007.
[54] J. Paris, F. Navarrina, I. Colominas, and M. Casteleiro, "Topology optimization of continuum structures with local and global stress constraints," *Structural and Multidisciplinary Optimization*, vol. 39, no. 4, pp. 419–437, 2009.
[55] A. Ramani, "Multi-material topology optimization with strength constraints," *Structural and Multidisciplinary Optimization*, vol. 43, pp. 597–615, 2011.
[56] S. Yamasaki, T. Nomura, A. Kawamoto, K. Sato, K. Izui, and S. Nishiwaki, "A level set based topology optimization method using the discretized signed distance function as the design variables," *Struct Multidisc Optim*, vol. 41, no. 5, pp. 685–698, Nov. 2009.
[57] R. A. Feijoo, A. A. Novotny, E. Taroco, and C. Padra, "The topological-shape sensitivity method in two-dimensional linear elasticity topology design," in *Applications of Computational Mechanics in Structures and Fluids*, CIMNE, 2005.
[58] G. Allaire and F. Jouve, "A level-set method for vibration and multiple loads structural optimization," *Structural and Design Optimization*, vol. 194, no. 30–33, pp. 3269–3290, 2005.
[59] I. Turevsky and K. Suresh, "Tracing the Envelope of the Objective-Space in Multi-Objective Topology Optimization," presented at the ASME IDETC/CIE Conference, Washington, DC, 2011.
[60] A. M. Mirzendehdel and K. Suresh, "A Pareto-Optimal Approach to Multimaterial Topology Optimization," *Journal of Mechanical Design*, vol. 137, no. 10, 2015.
[61] S. J. van den Boom, "Topology Optimisation Including Buckling Analysis," Delft University of Technology, Delft, 2014.
[62] K. Suresh, "A 199-line Matlab code for Pareto-optimal tracing in topology optimization," *Structural and Multidisciplinary Optimization*, vol. 42, no. 5, pp. 665–679, 2010.
[63] K. Suresh, "Efficient Generation of Large-Scale Pareto-Optimal Topologies," *Structural and Multidisciplinary Optimization*, vol. 47, no. 1, pp. 49–61, 2013.
[64] J. Céa, S. Garreau, P. Guillaume, and M. Masmoudi, "The shape and topological optimization connection," *Computer Methods in Applied Mechanics and Engineering*, vol. 188, no. 4, pp. 713–726, 2000.
[65] J. Hennessy and D. Patterson, *Computer Architecture, A Quantitative Approach*, 5th edutuin. Elsevier, 2011.
[66] K. Suresh and P. Yadav, "Large-Scale Modal Analysis on Multi-Core Architectures," in *Proceedings of the ASME 2012 International Design Engineering Technical Conferences & Computers and Information in Engineering Conference*, Chicago, IL, 2012.
[67] X. Bian, P. Yadav, and K. Suresh, "Assembly-Free Buckling Analysis for Topology Optimization," presented at the ASME-IDETC Conference, Boston, MA, 2015.
[68] Y. Jiang, H. Kautz, and B. Selman, "Solving problems with hard and soft constraints using a stochastic algorithm for MAX-SAT," in *Proceedings of the 1st International Workshop on Artificial Intelligence and Operations Research*, Timberline, Oregon, 1995.
[69] M. F. Ashby, "Multi-objective optimization in material design and selection," *Acta Materialia*, vol. 48, no. 1, pp. 359–369, 2000.
[70] M. Zhou and et. al., "Progress in Topology Optimization with Manufacturing Constraints," in *9th AIAA/ISSMO Symposium on Multidisciplinary Analysis and Optimization*, Georgia, Atlanta, 2002.
[71] P. Duysinx and O. Sigmund, "New developments in handling stress constraints in optimal material distribution," in *7th AIAA/USAF/NASA/ISSMO Symposium on Multidisciplinary Analysis and Optimization*, American Institute of Aeronautics and Astronautics, 2015.
[72] C. Le, J. A. Norato, T. E. Bruns, C. Ha, and D. A. Tortorelli, "Stress-based topology optimization for continua," *Structural and Multidisciplinary Optimization*, vol. 41, no. 4, pp. 605–620, 2010.
[73] L. A. Peterson and K. J. Londry, "Finite-Element Structural Analysis: A New Tool for Bicycle Frame Design. The Strain Energy Design Method," *Bike Tech, Bicycling Magazine's Newsletter for the Technical Enthusiast*, vol. 5, no. 2, 1986.
[74] R. R. Bini and et. al., "Pedal force effectiveness in Cycling: a review of constraints and training effects," *Journal of Science and Cycling*, vol. 2, no. 1, 2013.